\documentclass[12pt]{article}
\usepackage{epsfig}
\usepackage{amssymb}

\newcommand{\be}{\begin{equation}}
\newcommand{\ee}{\end{equation}}
\newcommand{\ba}{\begin{eqnarray}}
\newcommand{\ea}{\end{eqnarray}}
\newcommand{\bi}{\begin{itemize}}
\newcommand{\ei}{\end{itemize}}
\newcommand{\bn}{\begin{enumerate}}
\newcommand{\en}{\end{enumerate}}
\newcommand{\bc}{\begin{center}}
\newcommand{\ec}{\end{center}}

\def\unity{{\hbox{1\kern-.8mm l}}}

\newcommand{\gsim}{\lower.7ex\hbox{$\;\stackrel{\textstyle>}{\sim}\;$}}
\newcommand{\lsim}{\lower.7ex\hbox{$\;\stackrel{\textstyle<}{\sim}\;$}}

\def\mysection#1{\noindent {\bf #1} }

\begin{document}

\begin{titlepage}
\begin{flushright}
CERN-PH-TH/2006-159 \\  CPHT-RR-nnn.0806 \\ LPT-ORSAY-06-55 \\
{\tt hep-th/0608054}
\end{flushright}

\vskip.5cm
\begin{center}
{\huge \bf Dual realizations of dynamical symmetry breaking} \\
\vskip.1cm
\end{center}
\vskip0.2cm

\begin{center}
{\bf {Emilian Dudas}$^{a,b,c}$ and  {Chlo\'e Papineau}$^{c,b}$}
\end{center}
\vskip 8pt

\begin{center}
$^{a}$ {\it CERN Theory Division, CH-1211, Geneva 23, Switzerland} \\
\vspace*{0.1cm}
$^{b}$ {\it Centre de Physique Th\'eorique~\footnote{Unit{\'e} mixte du CNRS et de l'EP, UMR 7644.}, Ecole Polytechnique, 91128 Palaiseau Cedex, France} \\
\vspace*{0.1cm}
$^{c}$ {\it
 LPT~\footnote{Unit{\'e} mixte du CNRS, UMR 8627.},
B{\^a}t. 210, Univ. de Paris-Sud, 91405 Orsay Cedex, France} \\
\vspace*{0.3cm} {\tt Emilian.Dudas@cpht.polytechnique.fr,
chloe.papineau@th.u-psud.fr}
\end{center}

\vglue 0.3truecm

\begin{abstract}
\vskip 3pt \noindent We show the infrared equivalence between a recently proposed model containing a six dimensional scalar
field with a four-dimensional localized Higgs type potential and the
four-dimensional Nambu-Jona-Lasinio (NJL) model. In the dual NJL description, the fermions are localized at the origin
of a large two-dimensional compact space. Due to a  classical running effect above the compactification scale,
the four-fermion coupling of the NJL model increases from the cutoff scale down to the compactification scale, 
providing the large Fermi coupling needed for the dynamical symmetry breaking. We also present a string theory embedding of our field-theory construction. On more general grounds, our results suggest that 4d models with dynamical symmetry breaking can be given a higher
dimensional description in terms of field theories with nontrivial boundary conditions in the internal space.  

\end{abstract}

\end{titlepage}

\newpage


\section{Introduction and Conclusions}

Dimensional transmutation and generation of a small scale is a remarkable result common to many quantum
field theories, most notably the four-dimensional QCD and the two-dimensional Gross-Neveu model. The
effect is also realized in two-dimensional quantum mechanics with a deep (delta-like) attractive potential
and in six-dimensional scalar models\footnote{For gravitational aspects of codimension two models, see
e.g. \cite{6d}.} with 4d localized scalar potential and a large but compact
transverse space. The last example, put forward in \cite{dpr}, was analyzed from the point of the
quantum-mechanical problem in the case of a perturbative coupling $\mu$ appearing as the (dimensionless)
localized parameter interpreted as a mass term in 4d. It was shown that for a 6d scalar field with Dirichlet boundary condition
on a large two-dimensional compact space taken for simplicity to be a disk, there is a phase transition with a very light (compared
to the compactification scale) particle for a small critical value $\mu_c \ll 1$. The parameter $\mu$ was shown to
run between the cutoff scale $\Lambda$ and the compactification scale $R^{-1}$, such that precisely at the critical point, 
$\mu$ becomes large at $R^{-1}$. A very similar phenomenon of appearance of a light state close to a critical point where 
an (four-fermion) interaction becomes strong is in the
4d Nambu-Jona-Lasinio (NJL) model \cite{njl}.  The purpose of the present paper is to study closer this analogy and argue that
the 6d model studied in \cite{dpr} and the 4d NJL model are, in a sense which will be defined in detail later on, dual descriptions
of the same physics. 
Our starting point is to provide an explicit framework in which the 4d localized potential is generated,
by a Yukawa interaction of the bulk scalar field with N 4d localized fermions. In the large N limit,
integrating out the fermions produces precisely the potential needed for the symmetry breaking.
Alternatively, we show that integrating out the bulk scalar leads to a dual 4d NJL model with chiral symmetry breaking, where the
Fermi coupling is generated at tree-level by scalar bulk exchange. We show that the critical Fermi
coupling calculated by NJL methods in the large N limit agrees with the bulk 6d calculation of the
critical coupling calculated as a problem with nontrivial boundary conditions.     
   The 6d $\leftrightarrow$ 4d duality we study exchanges some quantum and classical natures of the symmetry breaking phenomenon.
In the bulk 6d picture, the quantum (Yukawa) interactions are completely encoded in a boundary condition, the localized 
scalar potential,  
whereas the symmetry breaking can be studied as a quantum-mechanical problem with nontrivial boundary condition and
can be understood as a result of a classical running effect in the transverse 2d space.  
In the 4d NJL picture, the symmetry breaking is provided by the nonperturbative self-consistent gap equation \cite{njl,nambu,bhl}, 
but in addition the four-fermion coupling has a classical logarithmic running between the cutoff and the compactification scale.
When in the bulk picture $\mu = \mu_c$ at $\Lambda$, the four-fermion coupling in the NJL picture at the compactification
scale reaches the critical value for the dynamical symmetry breaking $G (R^{-1}) = G_c$. 
  Our main interest in this equivalence is that, whereas a consistent treatment of the 
NJL model involves nonperturbative techniques like the large $N$ expansion or going below 4d and using UV nontrivial fixed points
and $1/\epsilon$ techniques, the bulk analysis is essentially classical\footnote{Throughout the paper by ``classical''
we mean classical from the point of view of quantum field theory, i.e. no quantum interaction. The treatment is still
quantum mechanical.} and does not need, in principle, any nonperturbative techniques.
 
The structure of the paper is as follows. In section 2 we review the six dimensional model worked out in \cite{dpr} and argue
that, in addition to the perturbative critical coupling $\mu_c$ we found there, there are other critical points corresponding
to large values $\mu_c^{(n)} > 1$. In section 2.1 we study a similar setup in which the nontrivial boundary condition in the compact
space is replaced by a bulk mass and Neumann boundary condition. In Section 3 we define the 6d model on an orbifold space instead of
a disk, which is more suitable for a microscopic (string theory) realization. Section 4 contains the main arguments concerning the
infrared equivalence between the 6d scalar model with nontrivial boundary condition and the 4d NJL model. Section 5 generalizes
the previous section to a (softly broken) supersymmetric theory. 
Section 6 provides an explicit string theory realization of the present setup in an orientifold of type IIB strings with D-branes.
\section{Six dimensional phase transition : perturbative and nonperturbative critical couplings}

Recently, \cite{dpr} addressed the problem of spontaneous symmetry breaking in a 6d scalar model 
with 4d localized Higgs potential. The corresponding action
reads\footnote{We are using a $(+,-,-,-,-,-)$ metric. The index $M$
denotes bulk coordinates and runs from $0,1,2,3,4,5$, while
$\mu=0,1,2,3$ denotes brane coordinates. We'll use either $x^{4,5}$
or $y_{1,2}$ to denote the two extra dimensions.} : 
\begin{eqnarray}
&& S \ = \ \int d^4 x d^2 y \ \biggl[ {1 \over 2} (\partial_M
\phi)^2 \ - \  V_\delta (\phi) \biggr] \ , \nonumber \\
&& V_\delta (\phi) \ = \ \left(-\frac{\mu^2}{2}\phi^2 +
\frac{\lambda}{4} \phi^4 \right) \cdot \delta^2 ({ y})  \ .
\label{npt1} 
\end{eqnarray}
 The scalar field $\phi$ has dimension two
and therefore $\mu^2$ is dimensionless. The scalar potential is
localized at the origin of the compact space. 
We resolve the singularity at ${ y}=0$ by introducing a disk $r < \epsilon$ supporting the potential,
\ba
&& V(\phi) \ = \ \frac{1}{\pi \epsilon^2}\left(-\frac{\mu^2}{2}\phi^2 +
\frac{\lambda}{4} \phi^4 \right)  \quad {\rm for} \ 0 < r < \epsilon \ ,  \label{npt01} \\
&& V(\phi) \ = \ 0  \quad \quad \quad \quad \quad \quad \quad \quad \quad \quad
{\rm for} \ \epsilon  < r < R \ .  \nonumber
\ea
According
to \cite{gw,dgv,dpr}, there is a "classical" running of the
tachyonic mass parameter 
\be 
{1 \over \mu^2 (Q)} = {1 \over \mu^2
(\Lambda)} - {1 \over 2 \pi} \ln ({\Lambda \over Q}) \ .
\label{pt18} 
\ee 
It was shown in
\cite{dpr} for a compact two-dimensional space, chosen to be a
disk of radius $R$, with 4d localized Higgs potential and Dirichlet boundary condition
\be
\phi |_{r =R} \ = \ 0 \ , \label{npt02}
\ee
that this model has a phase transition for a small
critical value 
\be {\mu_c^2 \over 2 \pi} \ln (R \Lambda) = 1 \ ,
\label{npt2} 
\ee 
where $\Lambda$ is a UV cutoff defined in connection with the resolution of the delta
singularity $\Lambda = 1 / \epsilon$. So the phase transition happens precisely when the renormalized value
$\mu_c^2 (R^{-1}) \rightarrow \infty$ blows up at the compactification scale. 
The running interpretation breaks down close to the
phase transition point. The classical running of $\mu$ induces also a running
for the self-coupling $\lambda$, according to the RG equation
\be
Q {d \lambda \over d Q} \ = \ - {2 \over \pi} \mu^2 \lambda \ , \label{pt018}
\ee
which, by using (\ref{pt18}), readily  integrates to
\be
\lambda (Q) \ = \ { \lambda (\Lambda) \over (1 - {\mu^2 \over 2 \pi} \ln {\Lambda \over Q})^4} 
\ . \label{pt019}
\ee
Notice that at the phase transition point $\mu = \mu_c$, 
\be
\mu (R^{-1}) \ \rightarrow \ \infty \quad , \quad \lambda (R^{-1}) \ \rightarrow \ \infty \ . \label{pt020} 
\ee

We will argue later on in section 4 that in a dual 4d theory which turns out to be a NJL theory, the conditions (\ref{pt020})
have the interpretation of compositeness conditions of \cite{bhl}. Close to the critical coupling, however, the running interpretation
breaks down and actually the higher-dimensional 6d and also the 4d physics turn out to be perturbative.
  
We now review and slightly update the arguments of \cite{dpr} by arguing that there are actually
additional but large critical couplings $\mu_c^{(n)} \ge \sqrt{4 \pi}$, 
defined by the presence of a 4d massless
mode in the spectrum. 
Assuming that this exists, slightly below it $\mu \le \mu_c$,  
in the background $\phi_c = 0$, the field eqs. for a 4d mode of mass $M^2 = p^2$, are  
\ba
&&  \Delta^{(2)} \phi \ + \ \frac{\mu^2}{\pi \epsilon^2} \phi=
0 \; , \;\;\; r < \epsilon \ , \nonumber \\
&&   \Delta^{(2)} \phi \ + p^2 \phi \ = \ 0 \; , \;\;\; r > \epsilon \ ,  \label{npt3}
\ea
where we neglected the mass $p^2$ inside the brane (this is a very good approximation for all masses much lighter than
the cutoff $\Lambda = 1/ \epsilon$). The solutions of (\ref{npt3}) with Dirichlet boundary condition 
(\ref{npt02}) and for $p^2 \ll R^{-2}$ are 
\ba
&& \phi (r) \ = \ f_0 \ J_0 ({\mu r \over \sqrt{\pi} \epsilon}) \quad , \qquad r < \epsilon \ , \nonumber \\ 
&&  \phi(r) \ = \ a \ \left[ \ln \frac{R}{r} - \frac{p^2 r^2}{4} \ln  \frac{R}{r} + \frac{p^2}{4}
(R^2 - r^2) \right]  \quad , \qquad  r > \epsilon \ .
\label{npt4}
\ea
The zero mass solutions $p^2=0$ define the whole set of critical couplings.
The matching conditions of the logarithmic derivative of the wave function at $r = \epsilon$ then give
\be
{\mu_c \ J'_0 ({\mu_c \over \sqrt{\pi}}) \over \sqrt{\pi} \ J_0 ({\mu_c \over \sqrt{\pi}})} \ \ln {R \over \epsilon}
\ = \ -1 \ . \label{npt5} 
\ee
For small $\mu_c$, eq. (\ref{npt5}) has the unique solution (\ref{npt2}), which is indeed small provided 
that $R^{-1} \ll \Lambda$. 
Equation (\ref{npt5}), however, has an {\it infinite discrete set} of solutions, as can easily be shown by a numerical
plot. The peculiarity of the perturbative solution (\ref{npt2}) is that the wave function  (\ref{npt4}) inside
the brane $r < \epsilon$, and actually also the wave functions of the massive modes below the cutoff $\Lambda$, are almost constant. 
On the contrary, the wavefunctions corresponding to the ``nonperturbative'' critical couplings 
(in the sense $\mu_c^2 / 4 \pi >1 $) 
have substantial variation inside the brane. Slightly below the critical coupling(s) $\mu_c$, the zero
mode becomes massive, with a mass $M$ given approximately by
\be
M^2 \ = \ p_{\mu} p^{\mu}  \ \simeq \ {2 (\mu_c^2 - \mu^2) \over \ \pi R^2} \  (\ln {R \over \epsilon})^2  \ 
\ . \label{npt6}  
\ee
The mass  (\ref{npt6}) reduces to the one computed in \cite{dpr} in the case of the perturbative critical
coupling (\ref{npt2}). 
Close to the other critical couplings $\mu \simeq \mu_c^{(n)}$ the light mode can also be described in a 4d effective field theory.
The 6d field is decomposed according to $\phi (x,r) = \sigma (x) \chi (r)$, where  
\ba
&& \chi (r) \ = \ \sqrt{2 \over \pi R^2} \ln {\left(R \over \epsilon\right)} \ \frac{J_0 ({\mu  r \over \sqrt{\pi} \epsilon})}{J_0({\mu \over \sqrt{\pi}})} \quad , \qquad r < \epsilon \ , \nonumber \\ 
&&  \chi(r) \ = \ \sqrt{2 \over \pi R^2}  \ \ln \frac{R}{r} \quad , \qquad  r > \epsilon \ .
\label{npt7}
\ea
The effective 4d potential for $\sigma$ is given by
\be
V_{eff} (\sigma) \ = \ {m_4^2 \over 2} \sigma^2 + {\lambda_4 \over 4} \sigma^4 \ , \label{npt8}
\ee
where the mass parameter and the coupling are given by
\ba
&& m_4^2 \ = \ - {2 \mu^2 \over \epsilon^2}  \int_0^{\epsilon} r dr \chi^2 + 
   {2 \mu^2 \over \epsilon^2}  \int_0^{\epsilon} r dr (\chi')^2  + 
2 \pi \int_{\epsilon}^R r dr  (\chi')^2 \ , \nonumber \\
&& \lambda_4 \ = \ {2  \lambda \over \epsilon^2} \ \int_0^{\epsilon} r dr \chi^4 (r) \ = \ 
{8 \lambda \over \pi^2 \mu^2 R^4} {1 \over J_0^4 ({\mu / \sqrt{\pi}})} \ (\ln \frac{R}{\epsilon})^4 \
\int_0^{\mu \over \sqrt{\pi}} x dx J_0^4 (x) \ , \label{npt9}
\ea
where the derivative in $\chi'$ is wrt the argument of the Bessel functions. Very close to $\mu_c^{(n)}$, the resulting mass
coincides with $M^2$ in (\ref{npt6}), showing the validity of the 4d description.   
The presence of light 4d modes close to the large critical couplings $\mu_c^{(n)}$ is a signature of a UV-IR mixing, where
the UV physics changes the masses in the IR. While in a microscopic theory in which $\mu$ is generated dynamically, large values
ask presumably for nonperturbative effects, from the bulk 2d viewpoint, $\mu$ changes only the boundary conditions of the
scalar field and its consequences can be treated exactly quantum-mechanically. On the other hand, the explicit values of $\mu_c^{(n)}$
depend on the way we regularize the origin of the 2d space and thus on the UV physics. Therefore, the physical
consequences of the large critical couplings are probably highly sensitive on the UV physics. This is not the case
for the small critical coupling (\ref{npt2}), whose value is insensitive to the regularization procedure and therefore
of the UV physics, as we explicitly check by using a different regularization in section 3.   
\subsection{Phase transition with bulk mass and boundary Higgs potential}

A natural question arising in the present setup is what happens if one replaces the positive contribution 
to the mass coming from the Dirichlet boundary condition by a bulk mass $m$, keeping the 4d localized Higgs-type
potential.  The action describing this case is given by
\begin{eqnarray}
&& S \ = \ \int d^4 x d^2 y \ \biggl[ {1 \over 2} (\partial_M
\phi)^2 \ - {1 \over 2} m^2 \phi^2 \ - \  V_\delta (\phi) \biggr] \ , \nonumber \\
&& V_\delta (\phi) \ = \ \left(-\frac{\mu^2}{2}\phi^2 +
\frac{\lambda}{4} \phi^4 \right) \cdot \delta^2 ({ y})  \ ,
\label{bulkmass1} 
\end{eqnarray}
where the field $\phi$ has now Neumann boundary condition
\be
\partial_r \phi |_{r = R} \ = \ 0 \ . \label{bulkmass2}
\ee
We are working in the unbroken phase $\phi_c = 0$, in which case the field equations for a 4d field of mass
$p^2$ are
\ba
&&  \Delta^{(2)} \phi \ + \ \frac{\mu^2}{\pi \epsilon^2} \phi=
0 \; , \;\;\; r < \epsilon \ , \nonumber \\
&&   \Delta^{(2)} \phi \ + (p^2-m^2) \phi \ = \ 0 \; , \;\;\; r > \epsilon \ .  \label{bulkmass3}
\ea
By defining $q^2 = m^2- p^2$, we find the solutions of (\ref{bulkmass3}) with Neumann boundary conditions (\ref{bulkmass2})
and for $q^2 \ll R^{-2}$ to be 
\ba
&& \phi (r) \ = \ f_0 \ J_0 ({\mu r \over \sqrt{\pi} \epsilon}) \quad , \qquad r < \epsilon \ , \nonumber \\ 
&&  \phi(r) \ = \ a \ \left[ 1 + \frac{q^2 R^2}{2} \ln  \frac{R}{r} + \frac{q^2}{4}
r^2 \right]  \quad , \qquad  r > \epsilon \ .
\label{bulkmass4}
\ea
Matching conditions at $r = \epsilon$ for zero mass solutions $p^2=0$ define the critical couplings in this case to be
given by the solutions of 
\be
\left[ 1 - {\mu_c \ J_1 ({\mu_c \over \sqrt{\pi}}) \over \sqrt{\pi} \ J_0 ({\mu_c \over \sqrt{\pi}})} \ 
\ln {R \over \epsilon} \right] \ {m^2 R^2 \over 2} \ = \ 
{\mu_c \ J_1 ({\mu_c \over \sqrt{\pi}}) \over \sqrt{\pi} \ J_0 ({\mu_c \over \sqrt{\pi}})} \ . \label{bulkmass5}
\ee
Analogously to the case discussed in the previous section, eq. (\ref{bulkmass5}) has an 
infinity but discrete number of solutions $\mu_c^{(n)}$, out of which only {\it one} is perturbative $\mu_c \ll 1$. 
In this perturbative case, similarly to (\ref{npt2}), there is a classical running interpretation of the critical coupling
\be
\pi R^2 m^2 \ = \ {\mu_c^2 \over 1 - {\mu_c^2 \over 2 \pi} \ln {R \over \epsilon}} \ = \ \mu_c^2 (R^{-1}) \ ,
\label{bulkmass6}
\ee  
where $ \mu_c^2 (R^{-1})$ is the renormalized value of the (perturbative) critical coupling at the compactification 
scale $Q = R^{-1}$. From  (\ref{bulkmass6}) it follows that for small bulk mass $m^2 R^2 \ll 1$, $\mu$ stays perturbative
at all energies above the compactification scale, whereas for large masses $m^2 R^2 \ge 1$, 
$\mu$ enters strong coupling regime if there is a light 4d mode in the spectrum.    

Slightly below the critical couplings, the light 4d mass is given by
\be
p^2 \ = \ m^2 \ - \ {2 \over R^2} \   {{\mu \ J_1 ({\mu \over \sqrt{\pi}}) / 
(\sqrt{\pi} \ J_0 ({\mu \over \sqrt{\pi}}))}
\over  1 - \ln {R \over \epsilon} {\mu \ J_1 ({\mu \over \sqrt{\pi}}) / (\sqrt{\pi} \ J_0 ({\mu \over \sqrt{\pi}}))} \ 
} \ . \label{bulkmass7}
\ee  
Very close to the perturbative critical coupling (\ref{bulkmass6}) , the light mass becomes
\be
p^2 \ = \ m^2 -  {1 \over \pi R^2} \ {\mu^2 \over 1 - {\mu^2 \over 2 \pi} \ln {R \over \epsilon}} 
 = \ m^2 -  {1 \over \pi R^2} \mu^2 (R^{-1}) \label{bulkmass8}
\ee
and has again a transparent interpretation in terms of the classical running between the compactification scale
and the cutoff. 
 
There are also light 4d states for  $q^2 \sim R^{-2}$ or larger. However, we are especially interested in 
the case of small bulk masses $m^2 R^2 \ll 1$, for reasons to be explained in the dual NJL 
formulation of a supersymmetric extension of this model.
\section{ Symmetry breaking phase transition in orbifolds}

More standard and easy to handle spaces in string theory are orbifolds. We will
consider in the following a compactification on the orbifold
$T^2/\mathbb{Z}_2$ and check as a warmup the properties of the phase
transition. The orbifold acts as the reflection $(y_1,y_2)
\rightarrow (-y_1,-y_2) $. This orbifold has four fixed points. The fixed points and their corresponding
$\mathbb{Z}_2$ coordinate transformations are summarized as:
\begin{equation}
     \label{orbifoldfix}
\begin{array}{cccc}
y_1 \to -y_1 ~& y_1 \to -y_1 + 2 \pi R_1~ &~ y_1 \to -y_1 ~&~
y_1 \to -y_1 + 2 \pi R_1 \\
y_2 \to -y_2 ~&~ y_2 \to -y_2  ~&~ y_2 \to -y_2 + 2 \pi R_2
~&~ y_2 \to -y_2 + 2 \pi R_2 \\
(0,0)~&~ (\pi R_1,0) ~&~ (0, \pi R_2)~&~ (\pi R_1, \pi R_2)
\label{pt2} \ .
\end{array}
\end{equation}
In complex notation, the action of $\mathbb{Z}_2$ on the compact space
is a  two-dimensional $\pi$ rotation,
$Z_2(y_1 +iy_2)$ = $e^{i \pi}(y_1 + i y_2)$.

The field equation is free in the bulk and has a delta function
source at the origin, suitably replaced by a mass distribution  
\be
    \label{sc1}
\partial_M \partial^M \Phi + {\partial V \over \partial \Phi} \, \delta^2 ({\bf y})= \ 0 \
. \label{pt3} \ee
Let us now proceed to study the mass spectrum for a scalar field
 with antiperiodic boundary conditions in the $y_1$-direction\footnote{This is the analog of the
Dirichlet boundary condition on the disk imposed in \cite{dpr}. As will become clear from our discussion, different boundary conditions, for
example $\Phi (y_1 ,y_2 + 2 \pi R_2) = - \Phi (y_1,y_2)$ or $\Phi (y_1 + 2 \pi R_1,y_2 + 2 \pi R_2) = - \Phi (y_1,y_2)$ lead to the same critical
coupling (\ref{npt2}). }
\be 
\Phi (y_1 + 2 \pi R_1,y_2) \ = \ - \ \Phi (y_1,y_2) \ .
\label{pt4} 
\ee

If the scalar field $\Phi$ is even under the orbifold action,
it can be decomposed on a complete basis formed by the cosine functions:
 \begin{equation}
\Phi(x,{\bf y})
 =
 \sum_{(k_1,k_2) \in \mathcal{I}}
 \frac{1}{\sqrt{2 \pi^2 R_1 R_2 }}
{\cos \left( \frac{k_1+1/2}{R_1} y_1 +  \frac{k_2}{R_2} y_2 \right)}
 \, \phi_{(k_1,k_2)} (x) \ . \label{pt5}
 \end{equation}
The indices $k_{1,2}$ belong to the set $\mathcal{I}$
\begin{equation}
    \label{pt6}
\mathcal{I} = \left\{ (0;0), (1\ldots \infty; 0), (0,-1;1\ldots
\infty), (1\ldots \infty; 1\ldots \infty), (1\ldots \infty; -\infty
\ldots -1) \right\}\ .
\end{equation}
In the unbroken vacuum, the quadratic part of the scalar action
takes the following form after integration over the two extra
dimensions
\begin{equation}
    \label{pt7}
\mathcal{L} = \mathcal{L}_{kin} - \frac{1}{2} \sum_{(k_1,k_2) \in
\mathcal{I}} \left( \frac{(k_1+1/2)^2}{R_1^2} + \frac{k_2^2}{R_2^2}
\right) \phi_{(k_1,k_2)}^2 + {\bar{\mu}^2} \left( \sum_{(k_1,k_2)
\in \mathcal{I}} \,  \, \phi_{(k_1,k_2)} \right)^2 \ ,
\end{equation}
where
\begin{equation}
\bar{\mu}^2 \equiv \frac{\mu^2}{4 \pi^2 R_1 R_2} \  \label{pt8}
\end{equation}
is the naive (volume suppressed) four dimensional lightest scalar
mass. The mass term of the 4d action  is
\begin{equation}
\mathcal{L}_{\rm mass} =
- {1 \over 2}
\sum_{(k_1,k_2),(p_1,p_2) \in \mathcal{I}}
 \phi_{(k_1,k_2)} \,
\mathcal{M}^2_{(k_1,k_2),(p_1,p_2)} \,
 \phi_{(p_1,p_2)} \ , \label{pt9}
\end{equation}
with the mass matrix given by
\begin{equation}
    \label{massmatrix1}
\mathcal{M}^2_{(k_1,k_2),(p_1,p_2)}  = - 2 {\bar \mu}^2 + \left(
\frac{(k_1+1/2)^2}{R_1^2} + \frac{k_2^2}{R_2^2} \right)
\delta_{k_1,p_1} \delta_{k_2,p_2} \ . \label{pt10}
\end{equation}
The diagonalization of this  mass matrix defines the physical mass
eigenstates.

Let us now try to find the eigenvalues and eigenvectors of the
mass matrix~(\ref{massmatrix1}). We use the techniques used in 5d models in \cite{ddg,bfz} and in 6d models in  \cite{dgv}.
The characteristic equation
is given
by
\begin{equation}
    \label{charectersticeq}
 \mathcal{M}^2 \Psi_m  \ = \  m^2 \Psi_m \ , \label{pt11}
\end{equation}
where $m^2$ represents the eigenvalues and $\Psi$ is the eigenvector
in the basis $\left| k_1,k_2 \right. \rangle_{(k_1,k_2) \in \mathcal{I}}$, i.e
$\Psi_{(k_1,k_2)} = \langle k_1, k_2 | \Psi_m \rangle$ .
The matrix equation~(\ref{charectersticeq}) is equivalent to the infinite set of explicit equations for every
$(k_1,k_2) \in \mathcal{I}$
\begin{equation}
2 \bar{\mu}^2 \Psi' = \left( - m^2 + \frac{(k_1+1/2)^2}{R_1^2} +
\frac{k_2^2}{R_2^2} \right) \Psi_{(k_1,k_2)} \ , \label{pt12}
\end{equation}
where $\Psi'$ is independent of $(k_1,k_2)$. 
The solution of the equations~(\ref{pt12}) is given by
\begin{equation}
\Psi_{(k_1,k_2)} \ =  \ \frac{\mathcal{N}}{ - m^2 + (k_1+1/2)^2/R_1^2 +
k_2^2/R_2^2}
 \ , \label{pt13}
\end{equation}
where $\mathcal{N}$ is a normalization constant independent of
$(k_1,k_2)$. Putting this solution back in the equation~(\ref{pt12})
and using the fact that 
\be 
\sum_{(k_1,k_2) \in \mathcal{I}} \ = \
{1 \over 2} \sum_{k_1,k_2 = \ -\infty}^{\infty} \ ,  
\ee
we obtain the eigenvalue equation
\begin{equation}
    \label{eigenvalue}
\frac{1}{\bar{\mu}^2} = \sum_{k_1, k_2 = -\infty}^\infty
\frac{1}{ - m^2 + (k_1+1/2)^2/R_1^2 + k_2^2/R_2^2 } \ , \label{pt14}
\end{equation}
or equivalently
\begin{eqnarray}
&& {1 \over \mu^2} \ = \ D (p^2 = m^2 , y_1=y_2=0) \quad , \quad {\rm where} \nonumber \\
&& D (p^2  , y_1,y_2) \ = \ {1 \over 4 \pi^2 R_1 R_2}  \sum_{k_1, k_2 = -\infty}^\infty
\frac{\cos [(k_1+1/2) y_1/R_1 + k_2 y_2/R_2] }{ - p^2 + (k_1+1/2)^2/R_1^2 + k_2^2/R_2^2 } \ \label{pt014} 
\end{eqnarray}
is the propagator in a mixed, 4d momentum and 2d position, representation.  
We want to find an estimate for the lightest solution, $m^2$, of the
eigenvalue equation (\ref{pt14}). The procedure we are using is
similar to the one used in \cite{dgv} and we only give the result
here. It is clear from (\ref{pt14}) that there is a critical
coupling, defined for arbitrary radii by 
\be 
{4 \pi^2 R_1 R_2 \over \mu_c^2} \ = \ \sum_{ |k_i| < k_i^{max}}
\frac{1}{(k_1+1/2)^2/R_1^2 + k_2^2/R_2^2 } \ , \label{pt15} 
\ee
which signals a second order phase transition, where the lightest
mass $m^2$ changes sign. For equal and large radii and by cutting the sums at $k_i^{max} = R \Lambda$, we find $\mu_c$
to be exactly equal to the value (\ref{npt2}) worked out in \cite{dpr}. 
A puzzle arises however in this KK approach to the phase transition. Indeed, whereas we accurately describe the perturbative critical
coupling (\ref{npt2}), eq. (\ref{pt15}) does not contain the nonperturbative couplings $\mu_c^{(n)}$ present in  (\ref{npt5}). We
believe that this is due to the way the logarithmic divergence is handled in (\ref{pt15}), or equivalently, to the ``brane
resolution'' for nonperturbative values of $\mu^2$. Indeed, as we already mentioned, in this case wave functions oscillate
significantly inside the brane and the regularization procedure becomes more subtle.  

Very close and slightly below the phase transition we can linearize
the mass equation (\ref{pt14}) in order to find for the lightest mode
\be
 m^2 \ \simeq \
{4 \pi^2 \over \alpha R_1 R_2} \ {\mu_c^2 - \mu^2 \over \mu_c^4}
\simeq \ {4 \pi^2 \over \alpha R_1 R_2} ( {1 \over \mu^2} - {1 \over
\mu_c^2}) \ ,
 \label{pt16}
 \ee 
where 
\be 
\alpha \ = \ R_1^{-2} R_2^{-2} \
\sum_{k_1, k_2 = -\infty}^\infty \frac{1}{[(k_1+1/2)^2/R_1^2 +
k_2^2/R_2^2]^2 } \ . 
\label{pt17} 
\ee 
The mass (\ref{pt16}) is not
exactly the same as the one worked out in \cite{dpr}. The reason is
that the geometries in \cite{dpr} and in the present section are
different and affect the IR physics, in particular physical masses. UV
physics, however is the same for both geometries ; in
particular the value (at the leading order in $\mu^2$) of the
perturbative critical coupling $\mu_c$ defining the phase transition is the same as in (\ref{npt2}). Since the regularizations
used on the disk and in (\ref{pt15}) are different, this shows the regularization independence of the perturbative critical 
coupling.   

Above the critical value, the scalar gets a vev (more precisely, a profile in the compact space) 
\be
\phi_c (y_1,y_2 ) \ = \ \phi_0 {\cal N} \sum_{k_1,k_2} {\cos [{(k_1+1/2) y_1 / R_1} + {k_2 y_2 / R_2}]
 \over  {(k_1+1/2)^2 / R_1^2} +  {k_2^2 / R_2^2}} \ , \label{pt015}
\ee
where according to \cite{dpr}
\be
\phi_0^2 \ = \ {\mu^2 - \mu_c^2 \over \lambda}  
\ee
and ${\cal N}$ is a normalization constant such that at the origin (more precisely, at the regularized mass distribution),
$\phi_c (0,0 ) = \phi_0$.  
\section{Localized matter and dual description : the NJL model}

An immediate question is how to generate in a natural way the
localized scalar potential needed for the symmetry breaking. The
simplest idea is to add (Weyl for definiteness, but the situation is
similar for Dirac fermions) $N$ fermions $\chi_i$ on the boundary,
with Yukawa couplings 
\be
g \chi_i \chi_i \phi (y=0) \ + \ {\rm h.c.} \ , \label{pt021}
\ee
to the (now complex) bulk field. The model has, in addition to a global $SO(N)$ symmetry, a continuous chiral $U(1)$
symmetry under which $\phi$ is charged. 

\subsection{The bulk picture}
One-loop quantum corrections generated by the fermion loops automatically
generate a scalar potential of the appropriate form (\ref{npt1}),
plus higher-order terms. The continuous chiral
symmetry will now be spontaneously broken at the phase transition.

In the large $N$ limit, the leading induced scalar potential is \be
V_{\rm eff} (\phi) \ = \ - \ N \int {d^4 p \over (2 \pi)^4} \  \ \ln
(p^2 + 4 g^2 |\phi|^2)  \ , \label{pt19}
 \ee
which can be expanded in powers of $\phi$ as
 \be
V_{\rm eff} (\phi) \ = \ - 4 N g^2 |\phi|^2 \int {d^4 p \over (2
\pi)^4} {1 \over p^2 } + 8 N  g^4 |\phi|^4 \int {d^4 p \over (2
\pi)^4} {1 \over p^4 } \ + \ \cdots \ . \label{pt20}
 \ee
We therefore induced radiatively, to the leading order in an $1/N$
expansion, a potential of the form (\ref{npt1}) with \be \mu^2 \ = \
{N g^2 \over 4 \pi^2} \ \ \Lambda^2  \ . \label{pt21} \ee As usual
\cite{cw} the power expansion in $\phi$ has severe IR divergences,
which are however resummed in the effective potential (\ref{pt19}).
Then the condition defining the symmetry breaking phase is \be {N
g^2 \over 4 \pi^2} \ \Lambda^2 \
> \ {4 \pi \over \ln (R^2 \Lambda^2)} \ , \label{pt22} \ee whereas
the perturbative expansion used in \cite{dpr}, for $\mu^2 \ll 1$,
translates here into \be {N g^2 \over 4 \pi^2} \ \Lambda^2  \ll 1 \ .
\label{pt23} \ee For $\langle \phi \rangle \not=0$ the brane
fermions $\chi_i$ acquire a mass and the chiral symmetry is
spontaneously broken, with the imaginary part of $\phi$ being the Goldstone boson.  
\subsection{Dual picture : the NJL model}
There is a dual description in which the bulk field $\phi$ is
integrated out at tree-level and the chiral symmetry breaking is
entirely described in terms of nonperturbative brane dynamics. The
resulting brane lagrangian has the simple form \be S_{\rm brane} \ =
- i \ \chi_i \sigma^{\mu} \partial_{\mu} {\bar \chi}_i
 \ + \ G \chi_i\chi_i {\bar \chi}_j {\bar \chi}_j \ ,
\label{pt24}
 \ee
 with
 \be
G \ = \ {g^2 \over 2 \pi^2 R_1 R_2} \sum_{(k_1,k_2) \in \mathcal{I}}
\  {1 \over \frac{(k_1+1/2)^2}{R_1^2} + \frac{k_2^2}{R_2^2}} \
\simeq {g^2 \over 4 \pi} \ \ln (R^2 \Lambda^2), \label{pt25}
 \ee
 where in the last equality we considered equal and large radii
 $R_1 = R_2 = R \gg M_*^{-1}$, where $M_*$ is the 6d fundamental (Planck) scale. Therefore the "dual"
 lagrangian (\ref{pt24}) is the Nambu-Jona-Lasinio model \cite{njl}, in which the chiral symmetry is dynamically
broken by the fermion condensate $\langle \chi_i \chi_i \rangle$
for values of the four-fermion coupling above the critical value
\be G \ > \ G_c \quad , \quad {\rm where} \ G_c^{-1} \ = \ {N
\Lambda^2 \over 4 \pi^2} \ \ . \label{pt26} \ee By using (\ref{pt21}) and
(\ref{pt25}), we find that the condition (\ref{pt26}) is precisely
the same  as the condition for the broken phase derived in the
"bulk" approach (\ref{pt22}).

\begin{figure}[htbp]
    \begin{center}
\centerline{
       \epsfig{file=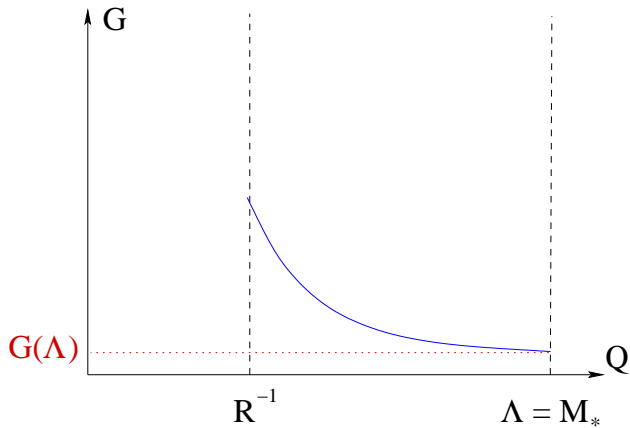,width=0.5\textwidth}
       }
          \caption{{\footnotesize
Classical running of the four-fermion coupling induced by the cumulative effect of the Kaluza-Klein states. The coupling
becomes strong $G=G_c$ at the compactification scale and induces chiral symmetry breaking.
}}
\label{fig:running}
    \end{center}
\end{figure}

Whereas in the deep IR the 6d bulk model is equivalent to the 4d NJL model, their UV behaviour is different\footnote{For earlier
ideas of the role of extra dimensions in dynamical symmetry breaking, see \cite{dobrescu}. For a recent extensive review on strong dynamics,
see e.g. \cite{hills}.}. In particular, due to the cumulative effects of the KK states, the four-fermion
coupling $G$ has a logarithmic running \be G (Q) \ \simeq {g^2 \over
4 \pi} \ \  \ln (\Lambda^2/Q^2) \label{pt27} \ee from the cutoff
scale $\Lambda$ to the compactification scale $R^{-1}$,
as illustrated in Fig.(\ref{fig:running}). So $G$
increases in the IR and can generate dynamical chiral symmetry
breaking. Even for couplings $g^2 \ll 4 \pi$ such that a perturbative treatment is available, the non-decoupling of heavy KK states 
generates a large four-fermion coupling in the infrared which drives the symmetry breaking. 

	\vspace{0.3cm}

In the bulk formulation, the mass parameter was
generated through quantum loops and the phase transition had a
"classical" (quantum mechanical) nature. In the NJL formulation, the
four-fermion coupling is generated classically while the phase
transition is generated in a nonperturbative fashion through the quantum gap equation 
\be 1 \ =
\ 4 N  G \int {d^4 p \over (2 \pi)^4} {1 \over p^2 + m^2} \ = \ 4 \ N \ G \ D (x=0,m^2)
 \ . \label{pt28} 
\ee 

In the previous expression, $D (x,m^2)$ is the 4d propagator in the position representation of a scalar field,  
and $m$ is the dynamical fermion mass. 

The counterpart of the NJL quantum gap
equation (\ref{pt28}) in the bulk formulation is the classical gap
equation (\ref{pt14})-(\ref{pt014}).  The similarity of the bulk equation (\ref{pt014}) and the NJL gap equation 
(\ref{pt28}) is transparent.  

Notice that the dual NJL description is valid in principle only below the phase transition. In this case and when the classical 
running interpetation is valid, the connection between the Higgs-localized 
scalar potential and the NJL model can also be understood in terms 
of the compositeness condition of \cite{bhl}. There, it was argued, by introducing a composite scalar field $H$, 
that the kinetic term $Z_H |\partial H|^2$ vanishes, $Z_H \rightarrow 0$, at the compositeness scale $\Lambda_C$. By a rescaling  
of the kinetic term to the canonical form, this is equivalent of imposing the boundary conditions $m_H \rightarrow \infty$, 
$\lambda_H \rightarrow \infty$,
where the parameters $m_H,\lambda_H$ are defined from the effective scalar potential $V = - m_H^2 |H^2| + \lambda_H |H|^4$.
In our 6d model, the running between the UV cutoff and the compactification scale $R^{-1}$ produces $\mu^2$ and $\lambda$ to
diverge at $R^{-1}$ precisely at the critical point. With a rescaling $\phi \rightarrow Z^{1/2} \phi$ of the scalar to go from the convention of \cite{dpr} to
the normalization of \cite{bhl}, keeping the scalar mass fixed, 
the wave function is
\be
Z (Q) \ = \ 1 - {\mu^2 \over 2 \pi} \ln {\Lambda \over Q} \ . \label{pt29}
\ee
At the phase transition $\mu = \mu_c$, we get
\be
Z (Q = R^{-1} , \mu=\mu_c) \ = \ 0 \quad , \quad \lambda (Q=R^{-1}) \ = \ \lambda (\Lambda) \ 
Z^{-2} (Q = R^{-1} , \mu=\mu_c) \ \rightarrow \infty \ .  \label{pt30}
\ee
The first of these conditions is similar to the one-loop (large N) induced wave function displayed
in \cite{bhl}, whereas the one for scalar self-coupling is different, since the UV physics is different. 
A natural candidate for the compositeness scale in our case is therefore $\Lambda_C = 1/R$.      

Analogously to the scalar parameters $\mu , \lambda$, the Yukawa coupling $g$ gets an induced running which can be easily
integrated :
\be
Q {d g \over d Q} \ = \ - {\mu^2 \over 2 \pi} g \quad , \quad g (Q) \ = \ 
{ g (\Lambda) \over 1 - {\mu^2 \over 2 \pi} \ln {\Lambda \over Q}} 
\ . \label{pt024}
\ee
The compositeness conditions have therefore to be supplemented with
\be
g (Q = R^{-1}) \ \rightarrow \ \infty \ , \label{pt025}
\ee
in analogy with the compositeness condition for the top quark in the top condensation scenario \cite{bhl}. 
Whereas the first and the third conditions (\ref{pt30})-(\ref{pt025}) are indeed similar to the compositeness conditions in \cite{bhl},
close to the critical coupling $\mu_c$ the running interpretation breaks down and the 4d physics is 
actually completely perturbative. Indeed,
defining the 4d effective theory as in \cite{dpr} by
\ba
&& \phi (x^{\mu},r) \ = \ \sqrt{2 \over \pi R^2} \ \ln {R \over \epsilon} \ \sigma (x^{\mu}) \quad , \quad r < \epsilon \ , \nonumber \\ 
&& \phi (x^{\mu},r) \ = \ \sqrt{2 \over \pi R^2} \ \ln {R \over r} \ \sigma (x^{\mu})   \quad , \quad \epsilon < r < R \ , \label{pt031} 
\ea
we find, at $\mu \simeq \mu_c$, the 4d scalar self-coupling and Yukawa coupling to the fermions to be
\be
\lambda_4 \ = \ {64 \pi^2 \over \mu_c^8} \ {\lambda \over R^4} \quad , \quad  g_4 \ = \ {2 \sqrt{2\pi} \over \mu_c^2 } \ {g \over R}
\ . \label{pt032}
\ee 
Since $1/\mu_c^2 \sim \ln (\Lambda R)$, the couplings (\ref{pt032}) are log enhanced compared to their naive (volume suppressed) values,
but are still perturbative and under control. 

Whereas the NJL 4d description and the bulk 6d one with appropriate boundary conditions are equivalent in the IR, the 6d picture can sometimes
be simpler to use in order to describe the symmetry breaking pattern. For example, consider two set of fermions living in two different 
fixed points,
$(0,0)$ and $(0,\pi R_2)$ and interacting both with (relatively large) Yukawa coupling to the same 6d bulk scalar, which has nontrivial boundary conditions     
(\ref{pt4}). In the NJL picture, there are three four-fermionic interactions coming from the two sets of fermions, with specific four-fermion
couplings. In the bulk picture, localized loops of fermions generate localized scalar potentials at both fixed points.
The scalar action in this case is
\begin{eqnarray}
&& S \ = \ \int d^4 x d^2 y \ \biggl[ {1 \over 2} (\partial_M
\phi)^2 \ - \  V_{\delta,1} (\phi) -   V_{\delta,2} (\phi) \biggr] \ , \nonumber \\
&& V_{\delta,1} (\phi) \ = \ \left(-\frac{\mu_1^2}{2}\phi^2 +
\frac{\lambda_1}{4} \phi^4 \right) \cdot \delta ({ y_1})  \delta ({ y_2})  \ , \nonumber \\
&& V_{\delta,2} (\phi) \ = \ \left(-\frac{\mu_2^2}{2}\phi^2 +
\frac{\lambda_2}{4} \phi^4 \right) \cdot \delta ({ y_1})  \delta ({ y_2 - \pi R_2})  \ . \label{pt31} 
\end{eqnarray}
It is a straightforward exercise to work out the equation defining the mass eigenstates (\ref{pt11}) in the unbroken phase. We find
\ba
&& \left[ {1 \over {\bar \mu}_1^2} -  \sum_{k_i = -\infty}^\infty \frac{1}{m^2 + (k_1+1/2)^2/R_1^2 + k_2^2/R_2^2 } \right]
\times \left[ {1 \over {\bar \mu}_2^2} -  \sum_{k_i = -\infty}^\infty \frac{1}{m^2 + (k_1+1/2)^2/R_1^2 + k_2^2/R_2^2 } \right] \nonumber \\
&& = \ \left( \sum_{k_i = -\infty}^\infty \frac{(-1)^{k_2}}{m^2 + (k_1+1/2)^2/R_1^2 + k_2^2/R_2^2 } \right)^2 \ . \label{pt32}
\ea   
Putting different fermions in different positions in the compact space can also provide a geometrical understanding of various values of their
Yukawa couplings via the wave function profile of the bulk scalar.

\section{Supersymmetric extension and coupling to localized chiral fields}

The basic mechanism we used for generating a sizable running uses the
logarithmic terms coming from the renormalization of the localized mass term. This implies
in particular that the corrections to the mass should be forbidden in
the bulk. The natural way to implement this is to have supersymmetry
in the bulk. Since the potential has to be generated on the brane,
we add boundary chiral fields in supersymmetric multiplets, with
supersymmetry softly broken on the boundary.

There are two inequivalent ways to supersymmetrize, by using either bulk hypermultiplets or bulk vector multiplets.   
Let us start  with the first case and consider a bulk hypermultiplet in 6d, which in 4d ${\cal N}=1$ supersymmetric
language has two chiral (super)fields $\Phi_{1,2} = (\phi_{1,2} ,
\psi_{1,2})$, and $N$ localized matter superfields $A_i = (z_i ,
\chi_i)$, with $i = 1 \cdots N $, where we denoted in parenthesis
the scalar and fermionic components of the multiplets. By using the
${\cal N}=1 $ superfield formalism of Ref. \cite{agw}, the action is
given by \ba  S \ = && \int d^2 y d^4 \theta \left( \Phi_1^{\dagger}
\Phi_1 + \Phi_2^{\dagger} \Phi_2 \right) + ( \int d^2 y d^2 \theta \
\Phi_1 \
(\partial_5 + i \partial_6) \ \Phi_2 + {\rm h.c.} ) \nonumber \\
&& + \int d^4 \ \theta A_i^{\dagger} A_i  (1 - \Sigma^2 \theta^2
{\bar \theta}^2)  + \left( \int d^2 \theta \ g \ \Phi_1 (y=0) \ A_i
A_i + {\rm h.c.} \right) \ , \label{se1} \ea

where $\Sigma$ is a scalar soft mass term for the boundary fields. In order to write the
component lagrangian, we first solve for the auxiliary fields \ba &&
F_{\phi,1}^{\dagger} \ = \ - (\partial_5 + i \partial_6) \phi_2 - g
z_i z_i
\delta^2 (y) \ , \nonumber \\
&&  F_{\phi,2}^{\dagger} \ = \  (\partial_5 + i \partial_6) \phi_1
\quad , \quad F_{A_i}^{\dagger}  = - 2 g \phi_1 z_i \ . \label{se2}
\ea

After eliminating auxiliary fields, the component lagrangian is \ba
S = &&\int d^2 y \left[ |\partial_M \phi_1|^2 + |\partial_M \phi_2|^2 - i
\psi_1 \sigma^{\mu} \partial_{\mu} {\bar \psi}_1 - i \psi_2
\sigma^{\mu} \partial_{\mu} {\bar \psi}_2 + (\psi_1 (\partial_5+i \partial_6) \psi_2 + {\rm h.c.}) 
\right] \nonumber \\
&& - \biggl[  S_{\rm kin} (z_i,\chi_i) + g \phi_1 \chi_i \chi_i + 2 g z_i
\psi_1 \chi_i + 4 g^2 |\phi_1|^2 |z_i|^2  \biggr. \nonumber \\
&& + g  \biggl. \left\{ z_i z_i (\partial_5 - i \partial_6) {\bar \phi}_2 +
{\rm h.c.} \right\} + \Sigma^2 |z_i|^2 + g^2 z_i^2 {\bar z}_j^2
\delta^2 (0) \biggr] \bigg|_{y=0} \ . \label{se3} \ea

Analogously to the non supersymmetric case we impose nontrivial
boundary conditions \be \Phi_{1,2} (y_1+ 2 \pi R_1, y_2) \ = \ - \
\Phi_{1,2} (y_1, y_2) \ . \label{se4} \ee  The $\mathbb{Z}_2$ orbifold has a
nontrivial action on the bulk hypermultiplet fields. A consistent
assignement is $\Phi_1$ to be even and $\Phi_2$ to be odd.
 The KK expansions in this case are
 \ba
&& \Phi_1 (x,{\bf y}) \ = \ \sum_{(k_1,k_2) \in \mathcal{I}}
 \frac{1}{\sqrt{2 \pi^2 R_1 R_2 }} \
{\cos \left( \frac{k_1+1/2}{R_1} y_1 +  \frac{k_2}{R_2} y_2 \right)}
 \, \Phi_1^{(k_1,k_2)} (x) \ , \nonumber \\
&& \Phi_2 (x,{\bf y}) \ = \ \sum_{(k_1,k_2) \in \mathcal{I}}
 \frac{1}{\sqrt{2 \pi^2 R_1 R_2 }} \
{\sin \left( \frac{k_1+1/2}{R_1} y_1 +  \frac{k_2}{R_2} y_2 \right)}
 \, \Phi_2^{(k_1,k_2)} (x) \ . \label{se5}
 \ea

The second possibility of supersymmetrization is to add bulk 6d vector multiplets, which in 4d ${\cal N} = 1 $ language are described
by vector $V$ and chiral $\phi$ supermultiplets, both in the adjoint representation of a gauge group $G$. 
In order to be able to couple $\phi$ to the localized matter, we need to choose $Z_2$, the orbifold action, such that $V$ is 
odd and therefore has no zero modes, whereas $\phi$ is even and can therefore couple to boundary chiral multiplets. 
More precisely, we can start from a nonabelian gauge group and give a nontrivial action of the orbifold on the gauge
degrees of freedom
\be
V (-y_1,-y_2) \ = \ P^{\dagger} \ V (y_1,y_2) \ P \quad , \quad  
\phi (-y_1,-y_2) \ = \ - \ P^{\dagger} \ \phi (y_1,y_2) \ P \ , \label{se05}
\ee
where $P$ is a matrix in the adjoint representation of the gauge group, such that $P^2 = 1$. The surviving (even) gauge group generators
$T_a $ satisfy $[T_a,P] = 0 $, whereas the remaining ones $T_{\alpha}$ are projected out. The complementary states 
$\phi_{\alpha} $ from the adjoint scalar
$\phi$ survive at the fixed points and can consistently be coupled to the localized matter. By imposing nontrivial 
boundary conditions on $\phi_{\alpha}$ we generate a setup where $\phi_{\alpha}$ plays the role of the scalar field with 
localized Higgs potential and can trigger the phase transition. While this can be an interesting alternative, we will not pursue 
this possibility further on.   

\subsection{Bulk picture}

The leading quantum corrections in $1/N$ come at one-loop with the
chiral (super)fields $A_i$ running in the loop. There is an
induced effective potential for $\phi_1$ which can be computed in
the standard way \cite{cw}. The result is \be V_{\rm eff} (\phi_1) \
= \ N \int {d^4 p \over (2 \pi)^4} \ \{ \ln (p^2 + \Sigma^2 + 4 g^2
|\phi_1|^2) - \ln (p^2 + 4 g^2 |\phi_1|^2) \} \ , \label{bp1}
 \ee
 which can be expanded in powers of $\phi_1$ as
 \be
V_{\rm eff} (\phi_1) \ = \ - 4 N \Sigma^2 g^2 |\phi_1|^2 \int {d^4 p
\over (2 \pi)^4} {1 \over p^2 (p^2+\Sigma^2)} + 8 N \Sigma^2 g^4
|\phi_1|^4 \int {d^4 p \over (2 \pi)^4} {2 p^2 + \Sigma^2 \over p^4
(p^2+\Sigma^2)^2} \ + \ \cdots \ . \label{bp2}
 \ee
We therefore induced radiatively, in the leading order in an $1/N$
expansion, a potential of the form (\ref{npt1}) with \be \mu^2 \ = \
{N g^2 \over 4 \pi^2} \ \Sigma^2 \ \ln {\Lambda^2 \over \Sigma^2} \
. \label{bp3} \ee In a first approximation, we can consider this as
the bare coupling in the model of \cite{dpr}. Then the condition
defining the symmetry breaking phase, for large and equal radii
$R_1=R_2 \gg {1 \over \Lambda}$, is \be {N g^2 \over 4 \pi^2} \ \Sigma^2 \ \ln
{\Lambda^2 \over \Sigma^2} \
> \ {4 \pi \over \ln (R^2 \Lambda^2)} \ , \label{bp4} \ee whereas
the perturbativity condition translates here into \be {N g^2 \over 4 \pi^2} \ \Sigma^2 \ \ln
{\Lambda^2 \over \Sigma^2} \ll 1 \ . \label{bp5} \ee For
$\langle \phi_1 \rangle \not=0$ the brane fermions $\chi_i$ acquire a
mass and the chiral symmetry is spontaneously broken. Finally, fermion loops also induce localized operators of the 
form $|(\partial_5 + i \partial_6) \phi_2|^2$. Their effect is to
renormalize the KK masses of the odd field $\phi_2$, but this effect has no relevance for our present discussion.


Notice that for natural values of (the dimensionful coupling) $g$, eq. (\ref{bp4}) can be satisfied only for large values 
of the soft breaking parameter $\Sigma$. This is easy to interpret in the dual NJL description, to which we now turn. 

\subsection{Dual description : the softly supersymmetric NJL model}

There is a dual description in which the bulk fields $\Phi_i$ are
integrated out at tree-level and the chiral symmetry breaking is
entirely described in terms of nonperturbative brane dynamics. Since
the bulk plus the interaction action is supersymmetric, the
integrating out procedure gives rise to a supersymmetric effective
action, to add to the brane lagrangian with softly broken
supersymmetry. There are some subtleties in proving that the
integration out leads to a well-defined four-dimensional
action without ill-defined (i.e.$\delta^2 (0)$) terms. Analogously
to former studies in 5d \cite{mp}, it can be checked that
the singular terms cancel out as they should. The resulting brane
lagrangian has the simple form 
\be 
S_{\rm brane} \ = \ \int d^4
\theta \{  A_i^{\dagger} A_i (1 - \Sigma^2 \theta^2 {\bar \theta}^2
) \ + \ G \ A_i A_i  A_j^{\dagger} A_j^{\dagger}  \} \ ,
\label{njl1}
\ee
 where
 \be
G \ = \ {g^2 \over 2 \pi^2 R_1 R_2} \sum_{(k_1,k_2) \in \mathcal{I}}
\  {1 \over \frac{(k_1+1/2)^2}{R_1^2} + \frac{k_2^2}{R_2^2}} \ .
\label{njl2}
 \ee
Therefore the "dual" lagrangian is a softly-broken  supersymmetric
version \cite{njlsusy1, njlsusy2} of the Nambu-Jona-Lasinio model
\cite{njl}. The dynamics of the softly broken supersymmetric version
of the NJL model in the large $N$ expansion was investigated in
detail in \cite{njlsusy2}. It was found there that chiral symmetry
is dynamically broken by the fermion condensate $\langle \chi_i
\chi_i \rangle$, for values of the four-fermion coupling above the
critical value \be G \ > \ G_c \quad , \quad {\rm where} \ G_c^{-1}
\ = \ {N \Sigma^2 \over 4 \pi^2} \ \ln {\Lambda^2 \over \Sigma^2} \
. \label{njl3} \ee
 
 By using (\ref{njl2}), we find that the condition
(\ref{njl3}) is precisely the same  as the condition for the broken
phase derived in the previous section, which for equal and large
radii is displayed in (\ref{bp4}). 

As in the 4d supersymmetric NJL model, there is a naturalness problem in the 6d construction.
In the 4d SUSY NJL model, the symmetry breaking occurs for
values of the four-fermion interaction $G \gsim 1/\Sigma^2 \gg 1/ \Lambda^2 $ much larger than its natural value. For small supersymmetry
breaking $\Sigma R \ll 1$, this generates strong four-fermion interactions well below the compositeness scale $\Lambda_c = R^{-1}$.
As transparent in (\ref{njl2}), the natural scale  of the strong four-fermion interactions for our 6d explicit realization 
is actually $R^{-1}$, unless $g$ is much larger than its natural value. The reason is simpler to understand
by rewriting (\ref{njl1}) in an appropriate form to compare with the Minimal Supersymmetric Standard Model (MSSM) \cite{njlsusy1}
\ba
S_{\rm brane} \ = && \ \int d^4
\theta \{  A_i^{\dagger} A_i (1 - \Sigma^2 \theta^2 {\bar \theta}^2) + H_1^{\dagger} H_1 \} \ + \nonumber \\
&& \ \left( \int d^2 \theta H_2 \ ( m H_1 - g \ A_i A_i ) \ + \ {\rm h.c.} \right)   \ . \label{njl4}
\ea        
Since $H_2$ is a Lagrange multiplier in (\ref{njl4}), the two lagrangians (\ref{njl1}) and (\ref{njl4}) are equivalent
for $G = g^2 / m^2$.  
As explained in \cite{njlsusy1,njlsusy2},  $H_2$ acquires a kinetic term $ Z_2 H_2^{\dagger} H_2$ at one-loop in the large $N$ expansion,
which vanishes at the compositeness scale $Z_2 (Q = \Lambda_c) \ = \ 0$. Below $\Lambda_c$, both $H_1$ and $H_2$ are dynamical fields
and are to be identified with the two Higgs doublets of the MSSM. In the language of (\ref{njl4}), the naturalness problem is that
in order to keep the Higgs mass light, the supersymmetric mass term should be of order $m \sim \Sigma$, which is nothing but the 
reincarnation of the so-called $\mu$-problem of MSSM
in the SUSY NJL case. In the 6d case, the analog of the action (\ref{njl4}) is (\ref{se1}), the analog of $H_2$ is $\Phi_1$, whereas
the analog of the supersymmetric mass $m$ is the mixing term $\Phi_1 (\partial_5+i \partial_6) \Phi_2$ in (\ref{se1}). Upon KK expansion, 
we get $m \sim (1 / 2R_1)$. Equivalently, this can be seen from the nontrivial boundary conditions (\ref{se4}) via the KK expansion (\ref{se5}).
Since the symmetry breaking only occurs for $m \sim \Sigma$, the model also requires large supersymmetry breaking scale, as already
anticipated. 

If we believe in the equivalence between the 6d scalar model and the 4d NJL model also close to the nonperturbative critical couplings
of section 2, then the corresponding Yukawa coupling $g$ can be large and produce a larger Fermi coupling $G$. However it will be hard to argue
reliably for very large Yukawas. 
Maybe a simpler road is to use a small bulk mass $m R \ll 1$, which will generate a large Fermi coupling
$G \sim g^2 / R^2 m^2$.      
Another solution in order to get symmetry breaking compatible with small supersymmetry breaking scale $\Sigma$ is to start with small
supersymmetry mass generated by boundary conditions. In other words, we need
\be
(\Phi_1 + i \Phi_2  ) (y_1 + 2 \pi R_1 , y_2) \ = \ e^{2 \pi i \omega} (\Phi_1 + i \Phi_2  ) (y_1 , y_2) \ , \label{njl5}
\ee     
with $\omega \ll 1$, in which case $m \sim (\omega /R_1)$. Whereas from first principles in string theory $\omega$ is quantized and cannot
be very small, in analogy with known 5d examples \cite{bfz}, $\omega \ll 1$ can actually be realized by starting
with periodic boundary conditions and adding small supersymmetric mass terms for bulk fields localized at the fixed points
$(\pi R_1,0)$ and/or $(\pi R_1, \pi R_2)$. Supersymmetry is broken softly with $\Sigma \Lambda \ll 1$ only at the origin
$(0,0)$. After re-diagonalization of the mass matrix, this is equivalent to starting with nontrivial
boundary conditions (\ref{njl5}) and no localized mass terms. This is technically natural in the sense that a small supersymmetric
mass term in the fixed points is protected by supersymmetry.

\section{String theory realization}

It is legitimate to ask if it is possible to realize the field theory construction we did provide in \cite{dpr} and
in this paper from a string theory framework. The answer is positive and we provide here one possible construction for the 
(softly broken) supersymmetric case of section 5\footnote{ Other realizations of 
dynamical symmetry breaking  can be found in \cite{harvey} for string realizations of a nonlocal NJL version and 
\cite{qcd} (\cite{alex})
for a string (field-theory) realization of the chiral symmetry breaking in QCD.}. The main requirement for the string construction is to
provide a large and flat two-dimensional compact space, so there should be no localized source which would curve the large 2d space. 
Indeed, our field theory analysis was done in flat space. 
This situation can be realized in orientifold constructions, where orientifold planes cancel the sources provided by the
branes. The basic ingredients of the construction are that the field $\phi$ arises from D$5$ branes wrapping our large 2d space,
whereas the localized fermions arise at the intersection between the D5 branes and a different set of branes, the intersection
being four dimensional and generating chiral fermions.  
We use orientifolds of type IIB string theory with D5-branes wrapping different coordinates of the internal space.
Our example is based on the orientifold projection $\Pi' = \Pi \ \pi_4 \pi_5 \pi_8 \pi_9$, where $\Pi$ is the left-right world-sheet interchange,
$\pi_4$, $\pi_5$ are parity  operations in two compact coordinates $x_4$ and $x_5$, to be identified with the 
two large dimensions $(y_1,y_2)$ in our field theoretical construction and $\pi_8$, $\pi_9$ are parity  operations in two internal
noncompact coordinates $x_8$ and $x_9$. The basic building block for brane configuration we consider is then the following
\ba
&& {\rm coord.} \quad \ 0 \qquad 1 \qquad 2 \qquad 3 \qquad 4 \qquad  5 \qquad  6 \qquad  7 \qquad  8 \qquad  9 
\qquad \ \nonumber \\
&& {\rm D}5_1 \  \qquad {\rm x} \qquad {\rm x} \qquad {\rm x} \qquad {\rm x} \qquad {\rm x} \qquad {\rm x} 
\qquad 0 \qquad 0 \qquad 0 \qquad 0 \nonumber \\
&& {\rm D}5_2  \  \qquad {\rm x} \qquad {\rm x} \qquad {\rm x} \qquad {\rm x} \qquad 0 \qquad 0
\qquad {\rm x}  \qquad {\rm x} \qquad 0 \qquad 0 \nonumber \\
&& {\rm O}5_2  \  \qquad {\rm x} \qquad {\rm x} \qquad {\rm x} \qquad {\rm x} \qquad 0 \qquad 0
\qquad {\rm x} \qquad {\rm x} \qquad 0 \qquad 0
\  \label{st1}
\ea  
In (\ref{st1}), crosses ${\rm x}$ denote coordinates parallel to the branes, whereas $0$ denotes orthogonal coordinates.
Notice that D$5_2$ branes and O$5_2$ planes are orthogonal to the 
large 2d space $(x_4,x_5)$. In order to keep the 2d space flat we 
need a configuration with D$5_2$ branes on top of the O$5_2$ planes, with locally zero tension and charge. 
We add a Wilson line $\langle W_4 \rangle$ on the D$5_1$ branes in the compact $x_4$ coordinate, which 
has the effect of breaking 
the gauge group and giving masses to fields charged under $W_4$. There are in particular
four charged scalar fields $\phi_i$, which get a mass from this Hosotani mechanism and will be identified with the master
field(s) of our field theory model. Notice first of
all that this field lives indeed in six dimensions, in the bulk of our large 2d compact space and it corresponds to a hypermultiplet
$\phi = (\phi_1, \phi_2)$ from the 4d viewpoint, as in section 5. The mass of $\phi$ 
is positive and it corresponds to the mass generated by boundary conditions
analyzed in \cite{dpr} and in sections 2, 4 and 5 of the present paper. Alternatively, using a Wilson line in one of the last 
four coordinates $x_6 \cdots x_9$ is equivalent to considering a bulk supersymmetric mass as in the section 3.  
For simplicity, we can consider the last three coordinates $x_7 \cdots x_9$ as being noncompact,
whereas $x_6$ is a circle and will be used to break supersymmetry a la Scherk-Schwarz. 

The D$5_2$ branes gauge fields are nondynamical in four dimensions and play the role of global symmetries.
The D$5_1$ brane degrees of freedom, on the other hand, are dynamical and contain in particular gauge fields and the field(s) $\phi$.
The D$5_1$-D$5_2$ sector, after additional orbifold projections to be discussed below, 
contains massless ${\cal N}=1$ chiral multiplets localized in  four dimensions, to be identified with the 4d chiral 
multiplets $A_i$ in section 5. 
At the effective low energy action level , the setup is similar to the one considered in section 5, with couplings of the 
form $A  \phi A $ and one expects the arguments presented there to apply
and generate dynamical symmetry breaking.   
  Non-trivial boundary conditions in the compact coordinate $x_6$ a la Scherk-Schwarz break supersymmetry at tree-level in the 
D$5_2$ sector, whereas the $D5_1$ branes, being orthogonal to the $x_6 $ coordinate, feel the breaking only through radiative
corrections \cite{ads1}. Notice that the dynamics in the large bulk coordinates $x_4,x_5$ is supersymmetric at that stage.
At one-loop, supersymmetry breaking propagates in the D$5_1$- D$5_2$ sector and generates
the localized 4d soft terms that were used in section 5. 

The setup preserve until now ${\cal N}=2$ supersymmetry in 4d spontaneously broken to ${\cal N}=0$ by the Scherk-Schwarz deformation, 
so additional ingredients are needed in order to generate chirality. The standard internal spaces used in this respect are the
Calabi-Yau spaces or the orbifolds. We choose here the second possibility. We introduce additional
$Z_2$ and $Z_3$ orbifold operations acting on the internal coordinates as 
\ba
&& Z_2 \ (z_1,z_2,z_3) \ = \  \ (- z_1,- z_2, z_3) \quad , \quad  Z_3 \ (z_1,z_2,z_3) \ = \  \ (e^{2 \pi i \over 3} z_1, z_2, 
e^{-{2 \pi i \over 3}} z_3) \ , \nonumber \\
&& {\rm where} \qquad z_1 \ = \ {x_4 + i x_5 \over \sqrt{2}} \quad , \quad   z_2 \ = \ {x_6 + i x_7 \over \sqrt{2}} \quad ,
\quad z_3 \ = \ {x_8 + i x_9 \over \sqrt{2}} \ \label{st2}
\ea
are the three complex internal coordinates. The resulting orientifold, which is dual to the so-called $Z_3 \times Z_2$ or 
$Z_6'$ type I orbifold in the literature \cite{afiv} after performing four T-dualities in $x_6,x_7,x_8,x_9$, 
reduces supersymmetry down to ${\cal N}=1$ in 4d. The 4d type I $Z_3 \times Z_2$ orbifold has D9 brane / O9 planes 
and one set of  D5 branes / O5 planes, wrapping the third internal torus. After the four T-dualities, the D9 branes (O9 planes)
become our D$5_1$ branes  (O$5_1$ planes) and the D5 (O5) branes become our  D$5_2$ branes  (O$5_2$ planes). Our Wilson line
is in the type I orbifold a Wilson line on the D9 branes.  

In order to break completely supersymmetry, as already announced we are adding a Scherk-Schwarz operation in the compact
coordinate $x_6$, compatible with the two orbifold operations. Our Scherk-Schwarz deformation
is a $2 \pi$ rotation in $x_6$ and one 4d spacetime coordinate. The corresponding worldsheet current anticommutes with the
$Z_2$ orbifold projection and commutes with the $Z_3$ one, as required by the consistency of the string construction \cite{ss}. 
A last subtlety, explained in detail in \cite{ads1} is that due to the Scherk-Schwarz operation, the  O$5_1$ planes, which are
perpendicular to the $x_6$ coordinate used in the supersymmetry breaking, are
actually pairs of  O$5_1$ orientifold - ${\overline O5}_1$ antiorientifold planes, situated at $x_6=0$ and $x_6 = \pi R_6$, respectively.
If the radius $R_6 \gg l_s$ is large enough, the closed string tachyon is massive and the timescale for the instability can 
be large enough. The D$5_1$ branes should be at (or close to) the point $x_6=0$, such that the strings  D$5_1$ -${\bar O5}_1$,
which break supersymmetry, to be very massive.  

A possible objection to the present setup is that the boundary conditions generated mass and/or the bulk hypermultiplet mass $m$ are 
not constant but field-dependent, given by the vev of the Wilson 
line on the D$5_1$  branes. Since the setup in non-supersymmetric, the Wilson line acquires a potential and its
vev will be dynamically fixed. This last point needs further investigation which is however beyond the goals of the present paper. 
This objection is also valid for the field-theory construction, in that we assumed that the radii of the large
2d space were stabilized by additional dynamics. 
To conclude, the setup presented in this section does realize the (softly) supersymmetric model work out in section 5, where the
supersymmetry breaking is soft and comes from a Scherk-Schwarz deformation in an extra space coordinate. 

The original non-superymmetric 
setup of \cite{dpr} and in sections 2,3 of the present paper can be in principle also realized for smaller values of 
$R_6 \gsim l_s$. However in this case
we expect severe tachyonic instabilities in the system, which need to be suppressed in order for the string picture to be a viable
description of the field-theory construction.     
\vskip 0.5cm

\mysection{Acknowledgements}: We thank T. Gherghetta, A. Pomarol and V.~Rubakov
for useful discussions. E.D. thanks the Galileo Institute of Theoretical Physics and INFN
for partial support during the completion of this work.
Work partially supported by the CNRS PICS \#~2530
and 3059, RTN contracts MRTN-CT-2004-005104 and MRTN-CT-2004-503369
and the European Union Excellence Grant, MEXT-CT-2003-509661.



\begin{thebibliography}{99}

\bibitem{6d}
The literature on gravitational aspects of codimension two models is vast. Some references are
A.~G.~Cohen and D.~B.~Kaplan,
  Phys.\ Lett.\ B {\bf 470} (1999) 52
  [arXiv:hep-th/9910132];
T.~Gherghetta and M.~E.~Shaposhnikov,
  Phys.\ Rev.\ Lett.\  {\bf 85} (2000) 240
  [arXiv:hep-th/0004014];
I.~Navarro,
  JCAP {\bf 0309} (2003) 004
  [arXiv:hep-th/0302129]; 
Y.~Aghababaie, C.~P.~Burgess, S.~L.~Parameswaran and F.~Quevedo,
  Nucl.\ Phys.\ B {\bf 680} (2004) 389
  [arXiv:hep-th/0304256];
G.~W.~Gibbons, R.~Guven and C.~N.~Pope,
  Phys.\ Lett.\ B {\bf 595} (2004) 498
  [arXiv:hep-th/0307238];
M.~Peloso, L.~Sorbo and G.~Tasinato
  Phys.\ Rev.\ D {\bf 73} (2006) 104025
  [arXiv:hep-th/0603026].

\bibitem{dpr}
  E.~Dudas, C.~Papineau and V.~A.~Rubakov,
  JHEP {\bf 0603} (2006) 085
  [arXiv:hep-th/0512276].
  
  
\bibitem{njl}
  Y.~Nambu and G.~Jona-Lasinio,
  Phys.\ Rev.\  {\bf 122} (1961) 345.
 
  
 \bibitem{nambu}
 Y.~Nambu,
EFI-88-62-CHICAGO
{\it Invited talk to appear in Proc. of 1988 Int. Workshop  New Trends in Strong Coupling 
Gauge Theories, Nagoya, Japan, Aug 24-27, 1988};
V.~A.~Miransky, M.~Tanabashi and K.~Yamawaki,
  Phys.\ Lett.\ B {\bf 221} (1989) 177.

\bibitem{bhl}
  W.~A.~Bardeen, C.~T.~Hill and M.~Lindner,
  Phys.\ Rev.\ D {\bf 41} (1990) 1647.


\bibitem{gw}
  W.~D.~Goldberger and M.~B.~Wise,
  Phys.\ Rev.\ D {\bf 65} (2002) 025011
  [arXiv:hep-th/0104170]; 
K.~A.~Milton, S.~D.~Odintsov and S.~Zerbini,
  Phys.\ Rev.\ D {\bf 65} (2002) 065012
  [arXiv:hep-th/0110051];
F.~del Aguila, M.~Perez-Victoria and J.~Santiago,
  JHEP {\bf 0302} (2003) 051
  [arXiv:hep-th/0302023].  

\bibitem{dgv}
  E.~Dudas, C.~Grojean and S.~K.~Vempati,
  arXiv:hep-ph/0511001.

  
\bibitem{ddg}
 K.~R.~Dienes, E.~Dudas and T.~Gherghetta,
  Phys.\ Rev.\ D {\bf 62} (2000) 105023
  [arXiv:hep-ph/9912455].
  
  
  
\bibitem{bfz}
J.~Bagger, F.~Feruglio and F.~Zwirner,
  JHEP {\bf 0202} (2002) 010
  [arXiv:hep-th/0108010].

 
\bibitem{cw}
  S.~R.~Coleman and E.~Weinberg,
  Phys.\ Rev.\ D {\bf 7} (1973) 1888.
  
  
  \bibitem{dobrescu}
  B.~A.~Dobrescu,
  Phys.\ Lett.\ B {\bf 461} (1999) 99
  [arXiv:hep-ph/9812349];
N.~Arkani-Hamed, H.~C.~Cheng, B.~A.~Dobrescu and L.~J.~Hall,
  Phys.\ Rev.\ D {\bf 62} (2000) 096006
  [arXiv:hep-ph/0006238];
M.~Hashimoto, M.~Tanabashi and K.~Yamawaki,
  Phys.\ Rev.\ D {\bf 64} (2001) 056003
  [arXiv:hep-ph/0010260].
V.~Gusynin, M.~Hashimoto, M.~Tanabashi and K.~Yamawaki,
  Phys.\ Rev.\ D {\bf 65} (2002) 116008
  [arXiv:hep-ph/0201106].
  

  
\bibitem{agw}
N.~Marcus, A.~Sagnotti and W.~Siegel,
  Nucl.\ Phys.\ B {\bf 224} (1983) 159;
  N.~Arkani-Hamed, T.~Gregoire and J.~G.~Wacker,
  JHEP {\bf 0203} (2002) 055
  [arXiv:hep-th/0101233].
  
  
  
\bibitem{mp}
  E.~A.~Mirabelli and M.~E.~Peskin,
  Phys.\ Rev.\ D {\bf 58} (1998) 065002
  [arXiv:hep-th/9712214];
E.~Dudas, T.~Gherghetta and S.~Groot Nibbelink,
  Phys.\ Rev.\ D {\bf 70} (2004) 086012
  [arXiv:hep-th/0404094].
  
  

  \bibitem{njlsusy1}
  W.~Buchmuller and S.~T.~Love,
  Nucl.\ Phys.\ B {\bf 204} (1982) 213;
 W.~Buchmuller and U.~Ellwanger,
  Nucl.\ Phys.\ B {\bf 245} (1984) 237.

\bibitem{njlsusy2}
T.~E.~Clark, S.~T.~Love and W.~A.~Bardeen,
  Phys.\ Lett.\ B {\bf 237} (1990) 235;
 M.~Carena, T.~E.~Clark, C.~E.~M.~Wagner, W.~A.~Bardeen and K.~Sasaki,
  Nucl.\ Phys.\ B {\bf 369} (1992) 33;
P.~Binetruy, E.~A.~Dudas and F.~Pillon,
  Nucl.\ Phys.\ B {\bf 415} (1994) 175
  [arXiv:hep-ph/9304278].


  
\bibitem{harvey}
E.~Antonyan, J.~A.~Harvey, S.~Jensen and D.~Kutasov,
  arXiv:hep-th/0604017;
A.~Parnachev and D.~A.~Sahakyan,
  arXiv:hep-th/0604173.
  
  
  
\bibitem{qcd}
 M.~Kruczenski, D.~Mateos, R.~C.~Myers and D.~J.~Winters,
  JHEP {\bf 0405} (2004) 041
  [arXiv:hep-th/0311270];
T.~Sakai and S.~Sugimoto,
  Prog.\ Theor.\ Phys.\  {\bf 113} (2005) 843
  [arXiv:hep-th/0412141].

\bibitem{alex}
J.~Erlich, E.~Katz, D.~T.~Son and M.~A.~Stephanov,
  Phys.\ Rev.\ Lett.\  {\bf 95} (2005) 261602
  [arXiv:hep-ph/0501128];
L.~Da Rold and A.~Pomarol,
  Nucl.\ Phys.\ B {\bf 721} (2005) 79
  [arXiv:hep-ph/0501218].
  
  

  
\bibitem{ads1}
I.~Antoniadis, E.~Dudas and A.~Sagnotti,
Nucl.\ Phys.\ B {\bf 544} (1999) 469 [arXiv:hep-th/9807011];
I.~Antoniadis, G.~D'Appollonio, E.~Dudas and A.~Sagnotti,
Nucl.\ Phys.\ B {\bf 553} (1999) 133 [arXiv:hep-th/9812118].



\bibitem{afiv}
G.~Zwart,
  Nucl.\ Phys.\ B {\bf 526} (1998) 378
  [arXiv:hep-th/9708040];
G.~Aldazabal, A.~Font, L.~E.~Ibanez and G.~Violero,
  Nucl.\ Phys.\ B {\bf 536} (1998) 29
  [arXiv:hep-th/9804026].
  
  
  \bibitem{ss}
R.~Rohm,
Nucl.\ Phys.\ B {\bf 237} (1984) 553;
S.~Ferrara, C.~Kounnas, M.~Porrati and F.~Zwirner,
Nucl.\ Phys.\ B {\bf 318} (1989) 75.



  
\bibitem{hills}
  C.~T.~Hill and E.~H.~Simmons,
  Phys.\ Rept.\  {\bf 381} (2003) 235
  [Erratum-ibid.\  {\bf 390} (2004) 553]
  [arXiv:hep-ph/0203079].
























  


\end{thebibliography}
\end{document}